# Simulated Quantum Computation of Molecular Energies

Alán Aspuru-Guzik*[a], Anthony D. Dutoi*[a], Peter J. Love[c] and Martin Head-Gordon[a,b]

[a] Department of Chemistry, University of California, Berkeley

[b] Chemical Sciences Division, Lawrence Berkeley National Laboratory

Berkeley, CA 94720 USA

[c] D-Wave Systems, Inc.

320-1985 West Broadway, Vancouver, BC V6J 4Y3 Canada

*These two authors contributed equally to this work.

# Abstract

abstract
The calculation time for the energy of atoms and molecules scales exponentially with system size on a classical computer, but polynomially using quantum algorithms. We demonstrate that such algorithms can be applied to problems of chemical interest using modest numbers of quantum bits. Calculations of the $H_2O$ and LiH molecular ground-state energies have been carried out on a quantum computer simulator using a recursive phase estimation algorithm. The recursive algorithm reduces the number of quantum bits required for the read-out register from approximately twenty to four. Mappings of the molecular wave function to the quantum bits are described. An adiabatic method for the preparation of a good approximate ground-state wave function is described and demonstrated for stretched $H_2$. The number of quantum bits required scales linearly with the number of basis functions used and the number of gates required grows polynomially with the number of quantum bits.

Feynman observed that simulation of quantum systems might be easier on computers using quantum bits (qubits) (1). The subsequent development of quantum algorithms has made this observation concrete (2-6). On classical computers, resource requirements for complete simulation of the time-independent Schrödinger equation scale exponentially with the number of atoms in a molecule, limiting such full configuration interaction (FCI) calculations to diatomic and triatomic molecules (7). Computational quantum chemistry is therefore based on approximate methods that often succeed in predicting chemical properties for larger systems, but their level of accuracy varies with the nature of the species, making more complete methods desirable (8).

Could quantum computation offer a new way forward for exact methods? Despite the formal promise, it has not been demonstrated that quantum algorithms can compute quantities of chemical importance for real molecular systems to the requisite accuracy. We address this issue by classically simulating quantum computations of the FCI ground-state energies of two small molecules. Although the basis sets used are small, the energies are obtained to the precision necessary for chemistry. Absolute molecular energies must be computed to a precision (greater than six decimal places) that reflects the smaller energy differences observed in chemical reactions (~0.1 kcal/mol). These simulations show that quantum computers of tens to hundreds of qubits can match and exceed the capabilities of classical FCI calculations.

A molecular ground-state energy is the lowest eigenvalue of a time-independent Schrödinger equation. The phase estimation algorithm (PEA) of Abrams and Lloyd (3,4) can be used to obtain eigenvalues of Hermitian operators; we address issues concerning its implementation for molecular Hamiltonians. The molecular ground-state wave function $|\Psi\rangle$ is represented on a qubit register **S** (state). Another register **R** (read-out) is used to store intermediate information and to obtain the Hamiltonian eigenvalue $E$. The Hamiltonian $\hat{H}$ is used to generate a unitary operator $\hat{U}$, with $E$ mapped to the phase of its eigenvalue $e^{i2\pi\phi}$.

$$\hat{U}|\Psi\rangle = e^{i\hat{H}\tau}|\Psi\rangle = e^{i2\pi\phi}|\Psi\rangle \qquad (1)$$
$$E = 2\pi\phi/\tau$$

Through repeated controlled action of powers of $\hat{U}$, the computer is placed in the state

$$|\mathbf{R}\rangle \otimes |\mathbf{S}\rangle = \left(\sum_n e^{(i2\pi\phi)n}|n\rangle\right) \otimes |\Psi\rangle \qquad (2)$$

The summation index $n$ enumerates the basis states of **R** according to their bit-string value. The quantum inverse Fourier transform is then applied to **R** to obtain an approximation to $\phi$ written in binary to **R**. The procedure is related to the Fourier transform of the time dependence of an eigenstate to obtain its eigenenergy. Using polynomially-scaling, classical approximation methods, an initial estimate of $E$ can be obtained, in order to choose $\tau$ such that $0 \leq (\phi \approx 1/2) < 1$.

We address four separate issues. First, we show how standard chemical basis sets can be used for representations of the wave function on **S**. Second, although the size of **R** relative to **S** will be marginal in the large-system limit, this initial overhead (20 qubits for a chemically meaningful result) presently represents a substantial impediment to both classical simulation and actual implementation of the algorithm. We show how a modification of the PEA makes it possible to perform a sequence of computations with a smaller register, such that the precision of the result obtained is independent of the size of **R**. Third, the algorithm requires that any estimated ground state has a large overlap with the actual eigenstate. We show how a good estimate of the ground-state wave function may be prepared adiabatically from a crude starting point. Finally, $\hat{U}$ must be represented in a number of quantum gates which scales polynomially with the size of the system, and we give such bounds.

Any implementation of a quantum simulation algorithm requires a mapping from the system wave function to the state of the qubits. Basis set methods of quantum chemistry often represent many-particle molecular wave functions in terms of single-particle atomic orbitals. The number of orbitals in a basis set is proportional to the number of atoms in a molecule. The molecular wave function may be represented by a state of **S** in two basic ways. In the *direct mapping*, each qubit represents the fermionic occupation state of a particular atomic orbital, occupied or not. In this approach, a Fock space of the molecular system is mapped onto the Hilbert space of the qubits. This mapping is the least efficient but has advantages discussed later. In the more efficient *compact mapping*, only a subspace of the Fock space with fixed electron number is mapped onto the qubits. The states of the simulated system and of the qubit system are simply enumerated and equated. Furthermore, one could choose only a subspace of the fixed-particle-number space. The compact mapping with restriction to a spin-state subspace is the most economical mapping considered in this work. Figure 1 shows that the number of qubits required for both the compact and direct mappings scales linearly with the number of basis functions. Also shown are the qubit requirements for specific molecules with different basis sets and mappings. More extensive qubit estimates for computations on $H_2O$ are given in Table 1, including restriction to the singlet-spin subspace.

In this work, a modified PEA was carried out, which uses a relatively small number of qubits in **R** (as few as four for stability). This implementation allows more of the qubits to be devoted to information about the system and decreases the number of consecutive coherent quantum gates necessary. This procedure can be interpreted as making continually better estimates of a reference energy. The Hamiltonian is then shifted by the current reference energy and an estimate of the deviation of the actual energy from the reference is computed. The reference energy is then updated and the procedure is repeated until the desired precision is obtained.

The algorithm at iteration $k$ is illustrated in Figure 2A. In iteration zero, we set $\hat{V}_0 = \hat{U}$ and perform a four-qubit PEA on $\hat{V}_0$. This estimates $\phi$ on the interval zero to unity with a precision of 1/16. We use this estimate to construct a shift $\phi_0$ which is a lower bound on $\phi$. We apply this shift and repeat the four-qubit PEA using the new operator $\hat{V}_1 = \left[ e^{-i2\pi\phi_0} \hat{V}_0 \right]^2$. This determines the remainder of $\phi$ above the previous lower bound on an interval representing half of the previous interval. In each subsequent iteration $k$, we construct a similarly modified operator $\hat{V}_k$ and shift $\phi_k$. By choosing a $\phi_k$ which is 1/4 lower than the phase of the $\hat{V}_k$ eigenvalue estimate, we ensure that the phase of the $\hat{V}_{k+1}$ eigenvalue is approximately centered on the interval zero to unity. In each iteration, we therefore obtain one additional bit of $\phi$, as shown in Figure 2B for a calculation on $H_2O$.

To demonstrate the usefulness of the recursive procedure, we carried out calculations on $H_2O$ and LiH. For $H_2O$, we used the minimal STO-3G basis set, yielding 196 singlet-spin configurations; there are 1210 such configurations for LiH in the larger 6-31G basis. This required eight and eleven qubits, respectively, for the compact mapping of the singlet subspace. Register **S** was initialized to the Hartree-Fock (HF) wave function in both cases. After 20 iterations, the electronic energy obtained for $H_2O$ (-84.203663 a.u.) matched the Hamiltonian diagonalization energy (-84.203665 a.u.). The LiH calculation (-9.1228936 a.u.) matched diagonalization (-9.1228934 a.u.) to the same number of significant digits. The precision is good enough for almost all chemical purposes. The discrepancy between the PEA and diagonalization is attributed to error in matrix exponentiation to form $\hat{U}$ from $\hat{H}$.

In the simulations described above, the approximation to the ground-state wave function was the HF state $|\Psi^{HF}\rangle$. The probability of observing the exact ground state $|\Psi\rangle$, and hence the success of the PEA, is then proportional to $|\langle\Psi|\Psi^{HF}\rangle|^2$. However, it is known for some cases, such as molecules close to the dissociation limit or in the limit of large system size, that the HF wave function has vanishing overlap with the ground state (9). The overlap of the initially prepared state with the exact state can be systematically improved by an adiabatic state preparation (ASP) algorithm, relying on the adiabatic theorem (10-12). The theorem states that a system will remain in its ground state if the Hamiltonian is changed slowly enough. Our Hamiltonian is changed slowly by

discretized linear interpolation from the trivial HF case to the FCI operator. The efficiency is governed by how rapidly the Hamiltonian may be varied, which is determined by the gap between ground-state and first-excited-state energies along the path (11). In the case of quantum chemistry problems, lower bounds on this gap may be estimated using conventional methods.

The path $\hat{H}^{HF} \to \hat{H}$ is chosen by defining $\hat{H}^{HF}$ to have all matrix elements equal to zero, except the first element, namely $H_{1,1}$, which is equal to the HF energy. This yields an initial gap the size of the ground state mean-field energy, which is very large relative to typical electronic excitations. The ASP method was applied to the $H_2$ molecule at large separations in the STO-3G basis, for which the squared overlap of the HF wave function with the exact ground state is one half. As evidenced by Figure 3A, the ASP algorithm prepares states with a high squared overlap for several internuclear distances of the $H_2$ molecule. Figure 3B plots the relevant gap along the adiabatic path, which is shown for this system to be well-behaved and non-vanishing.

The accuracy and quantum-gate complexity of the algorithm depend on the specific gate decomposition of the unitary operators $\hat{V}_k$, defined above. The factorization of unitary matrices into products of one- and two-qubit elementary gates is the fundamental problem of quantum circuit design. We now demonstrate that the length of the gate sequences involved are bounded from above by a polynomial function of the number of qubits.

We analyze the gate complexity of our $\hat{U}$ for the direct mapping of the state. The molecular Hamiltonian is written in second quantized form as

$$\hat{H} = \sum_X \hat{h}_X = \sum_{p,q} \langle p|\hat{T}+\hat{V}_N|q\rangle \hat{a}_p^+\hat{a}_q - \frac{1}{2}\sum_{p,q,r,s}\langle p|\langle q|\hat{V}_e|r\rangle|s\rangle \hat{a}_p^+\hat{a}_q^+\hat{a}_r\hat{a}_s \qquad (3)$$

where $|p\rangle$ is a one-particle state and $\hat{a}_p$ is its fermionic annihilation operator; $\hat{T}$, $\hat{V}_N$, and $\hat{V}_e$ are the one-particle kinetic and nuclear-attraction operators and the two-particle electron-repulsion operator, respectively. It has been shown in (2), that for the following approximation to $\hat{U}$,

$$e^{i\hat{H}\tau} \approx \left[\prod_X e^{i\hat{h}_X \frac{\tau}{M}}\right]^M \qquad (4)$$

$M$ can always be chosen such that the error is bounded by some preset threshold. The number of gates to implement $\hat{U}$ then scales polynomially with the system size for a given $M$, under the condition that the number of terms $\hat{h}_X$ scales polynomially with system size, and that each $\hat{h}_X$ acts on a fixed maximum number of qubits. In our case, these conditions are manifestly fulfilled. The number of terms in the Hamiltonian grows approximately with the fourth power of the number of atoms, and each term involves maximally four basis functions, implying action on at most five qubits in the direct mapping (four qubits in **S** plus a control qubit in **R**). $M$ is a multiplicative factor in the number of gates. Since the fraction of all pairs of $\hat{h}_X$ terms that do not commute decreases with system size, it is reasonable to assume that $M$ increases polynomially at worst. The advantage of the direct mapping is that, at most, controlled four-qubit unitary operations are required. The number of one- and two-qubit elementary gates required to represent an arbitrary four-qubit gate has been shown to be always less than 400 (13); the structure of a controlled four-qubit unitary will allow a decomposition into a similar order of magnitude in the number of gates.

We have found that chemical precision can be achieved with modest qubit requirements for the representation of the wave function and for the read-out register. The ASP algorithm has been shown to systematically improve the probability of success of the PEA. While exponentially difficult on a classical computer, extension to larger molecules requires only linear growth in the number of qubits. The direct mapping for the molecular wave function to the qubit state allows the unitary operator to be decomposed into a number of gates which manifestly scales polynomially with system size.

The difficulty of performing quantum-computing simulations is approximately an order of magnitude greater than conventional FCI. While possible as experiments, such simulations are not a competitive alternative. To repeat the calculations performed here with a high-quality basis set (cc-pVTZ) would require **S** to consist of 47 or 22 qubits for $H_2O$ or LiH, respectively, using the compact mapping of the full Hilbert space. For most molecules and basis set combinations shown in Figure 1, a FCI calculation is certainly classically intractable. A FCI calculation for $H_2O$ with cc-pVTZ would be at the edge of what is presently possible. This demonstrates an often-stated conjecture, that quantum simulation algorithms with 30 to 100 qubits will be among the smallest applications of quantum computing that can exceed the limitations of classical computing.

Table 1

| water | Basis set (number of functions) | | |
|---|---|---|---|
| Mapping | STO-3G (7) | 6-31G* (19) | cc-pVTZ (58) |
| compact (singlets) | 8 | 25 | 42 |
| compact | 10 | 29 | 47 |
| direct | 14 | 38 | 116 |

Table 1 [Qubit requirements for computations on water]: The number of qubits needed to store the wave function of water is given for various basis sets and system-qubit mappings, including restriction to the singlet-spin subspace.

Figure 1

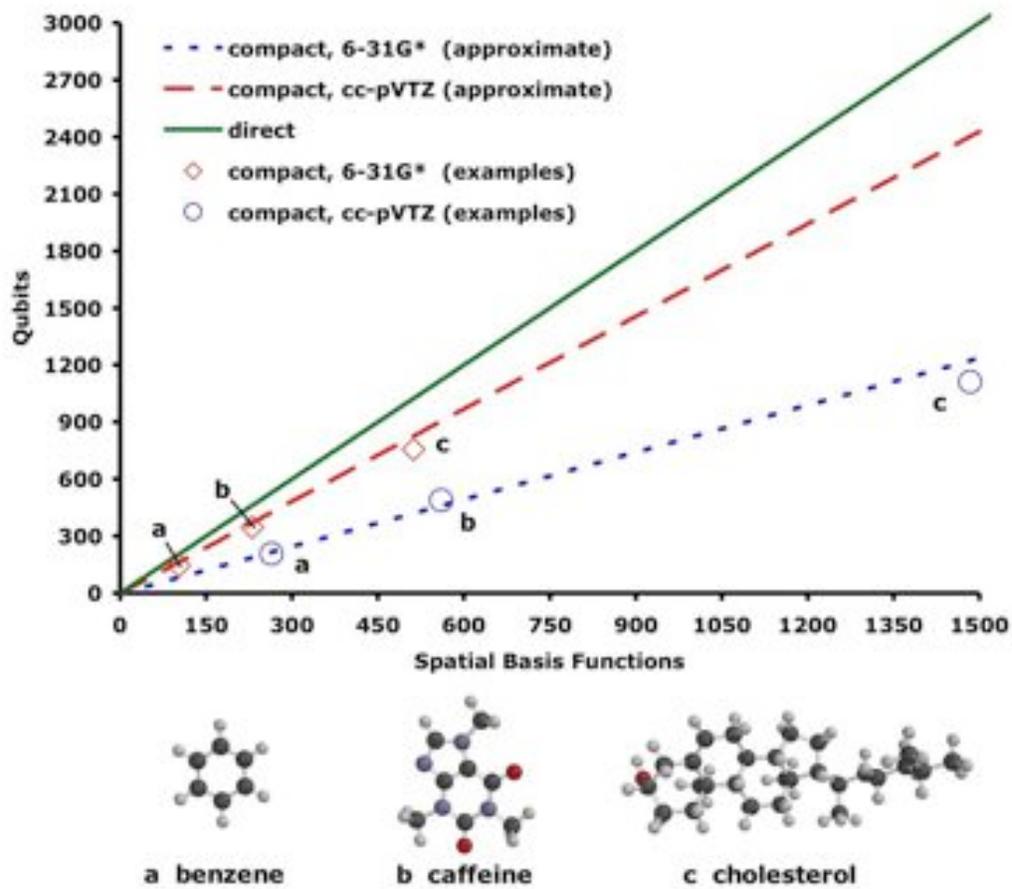

a benzene    b caffeine    c cholesterol

Figure 1 [Qubit requirements vs. Basis size]: The number of qubits required to store the wave function of a molecule is shown as a function of the number of basis functions for two different mappings. For the compact mapping, the qubit requirement depends also on the ratio of number of electrons to basis functions, which is relatively constant for a given basis set; although the higher-quality cc-pVTZ basis is more economical per basis function, a molecule in this basis uses substantially more functions than with the 6-31G* basis. The qubits required for specific molecules and basis sets are also shown.

Figure 2

A

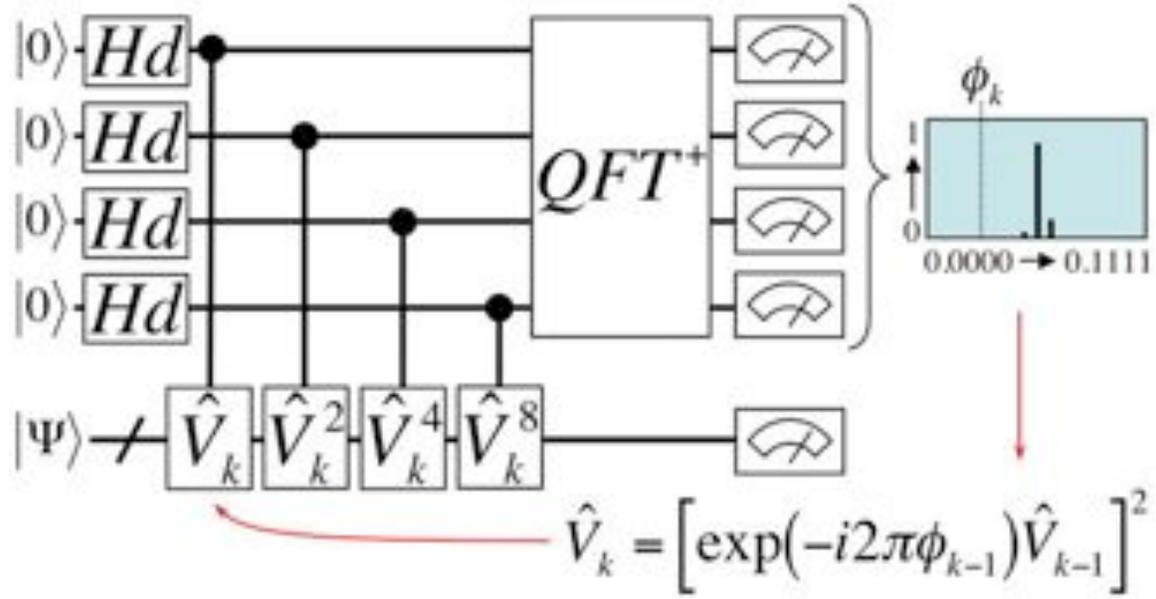

B

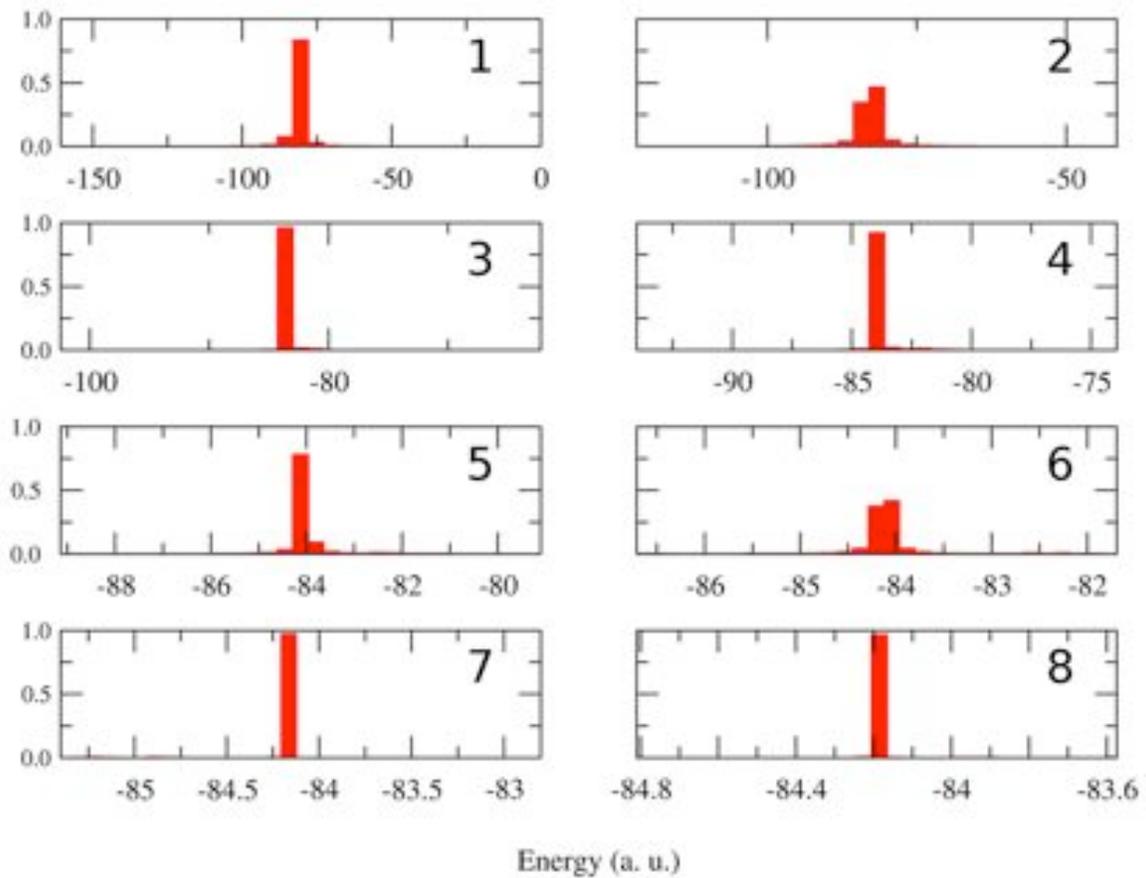

Energy (a. u.)

Figure 2 [Recursive PEA circuit and output]: (A) The quantum circuit for the recursive phase estimation algorithm is illustrated. $k$ iterations are required to obtain $k$ bits of a phase $\phi$ that represents the molecular energy. $QFT^+$ represents the quantum inverse Fourier transform and $Hd$ is a Hadamard gate; the dial symbols represent measurement. (B) Output probabilities for obtaining the first eight bits of $\phi$ in the water calculation are shown. The abscissa is scaled to be in terms of molecular energy and the ordinate is probability.

Figure 3

A

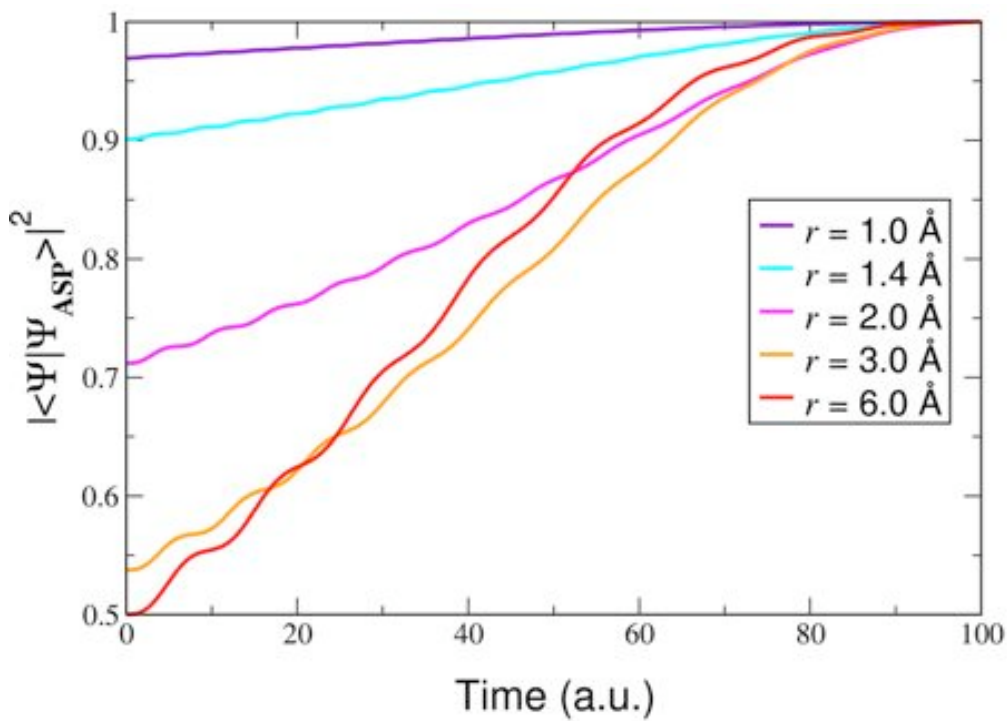

B

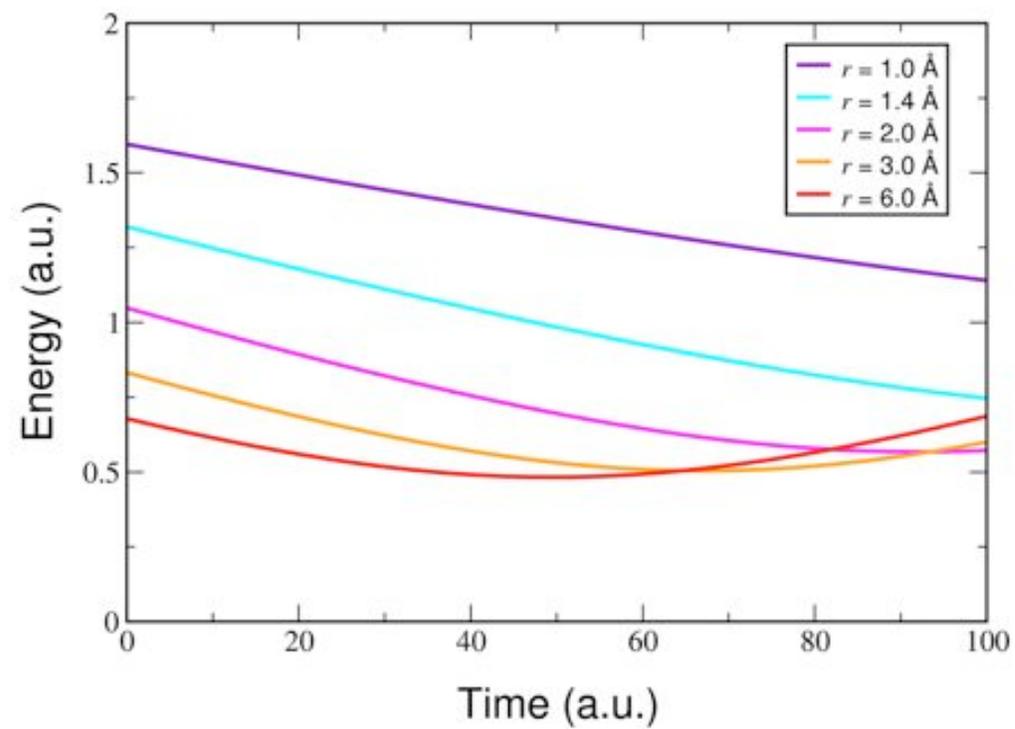

Figure 3 [ASP evolution of ground-state overlap and excitation gap]: (A) Time evolution of the squared overlap of the wave function $|\Psi_{ASP}\rangle$ with the exact ground state $|\Psi\rangle$ during adiabatic state preparation is shown. The system is the hydrogen molecule at different nuclear separations $r$; time was divided into 1000 steps in all cases. (B) Time evolution of the ground- to first-excited-state energy gap of the Hamiltonian used along the adiabatic path is shown.

## Methods

The calculation for the water molecule was performed using the experimental geometry (r= 95.76 pm, ∠=104.51°) (*S1*). The LiH molecule was calculated at r=140 pm. The basis sets used were STO-3G for water (*S2*) and 6-31G for LiH (*S3*). The Hamiltonian matrix was obtained using a modified version of the Q-Chem code (*S4*). Numerical exponentiation was carried out using the EXPOKIT package (*S5*). Quantum simulation was performed with the TeQUiLA library (*S6*).